\documentclass[journal=jpclcd, manuscript=article]{achemso}
\usepackage{amsmath}
\author{Michael Sch\"uler}
\email{michael.schueler@physik.uni-halle.de}
\author{Yaroslav Pavlyukh}
\author{Jamal Berakdar}
\affiliation{Institute for Physics, Martin-Luther University Halle-Wittenberg \\
  Heinrich-Damerow-Stra\ss{}e 4, 06120 Halle}

\title{Local Ionization Dynamics Traced by Photo-assisted Scanning
  Tunneling Microscopy: a Theoretical Approach}

\begin{document}

%%%%%%%%%%%%%%%%%% title page information %%%%%%%%%%%%%%%%%%

\begin{abstract}
 For tracing the spatiotemporal evolution
of electronic systems we suggest and analyze theoretically a  setup that exploits  the
excellent spatial resolution based on scanning tunneling microscopy techniques
 combined with
 the temporal resolution of femtosecond pump-probe photoelectron spectroscopy.
  As an example we consider the
laser-induced,  local vibrational dynamics of a
surface-adsorbed molecule. The photoelectrons released by a  laser pulse can be
collected by the scanning tip and utilized to access the spatio-temporal
dynamics. Our proof-of-principle calculations are based on the solution of the
time-dependent Schr\"odinger equation  supported by the \emph{ab initio}
computation of the matrix elements determining the
dynamics.
\begin{center}
\includegraphics[width=7cm,angle=-90]{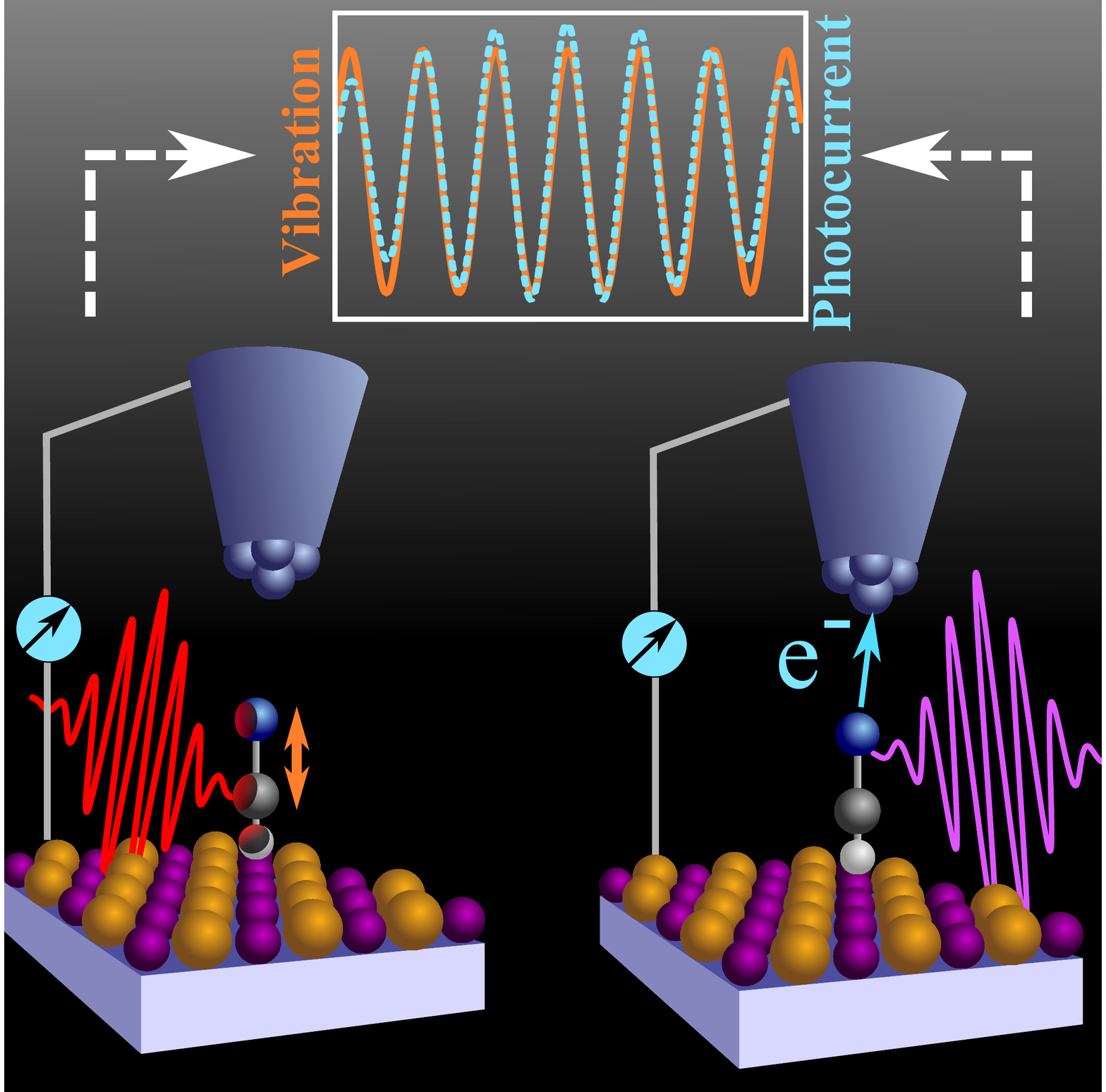}
\end{center}
\rule{\linewidth}{0.5mm}
\textbf{Keywords}: surface photochemistry, surface vibrations, time-resolved photoemission, photo-assisted STM, Dyson orbital, orbital imaging
\end{abstract}

%%%%%%%%%%%%%%%%%%%%%%%%%%  body  %%%%%%%%%%%%%%%%%%%%%%%%%%
The impressive advances made in generating and utilizing sub-femtosecond
laser pulses
\cite{goulielmakis_single-cycle_2008,schultze_delay_2010,haessler_attosecond_2010,
  shafir_atomic_2009,krausz_attosecond_2009,cavalieri_attosecond_2007}
for time-resolving  the dynamics of electronic systems
are paralleled with the fascinating versatility of scanning tunneling
microscopy (STM) \cite{oka_spin-dependent_2010,wiesendanger_spin_2009,fu_reversible_2012,schwobel_real-space_2012}
that allows for atomic spatial resolution.
 STM  accesses  also spectroscopic information that are related, under certain assumptions, to the
 local density of states of the probe. The transport through
 single molecules
\cite{mullegger_spectroscopic_2009,katano_single-molecule_2010,burema_resonance_2012,jiang_observation_2012,zhao_superatom_2009,jorn_current_2011}
provides, for instance, insight in the analysis and manipulation of the molecular properties like the
conformation \cite{braun_probing_2005,alemani_electric_2006,kuck_naked_2008},
or the characterization of magnetic systems \cite{loth_measurement_2010,loth_bistability_2012,robles_spin_2012}.

To explore the sub-pico second local  dynamics, there have been proposals for inducing and probing  the electron
dynamics by means of focused femtosecond laser pulses applied within the STM context
\cite{pfeiffer_rapid_1997,gerstner_femtosecond_2000,grafstrom_photoassisted_2002,lee_laser_2010}.
The experimental limitations and issues experienced by earlier attempts
(mainly due the thermal expansion of the tip)
\cite{pfeiffer_rapid_1997,gerstner_femtosecond_2000,grafstrom_photoassisted_2002}
have been shown to be possible to be circumenvented by ultrafast two-photon schemes\cite{lee_laser_2010,dolocan_two-color_2011}.

\begin{figure}[ht!]
\centering
\rule{\linewidth}{0.4mm}\vspace{0.5cm}
\includegraphics[height=13cm,angle=-90]{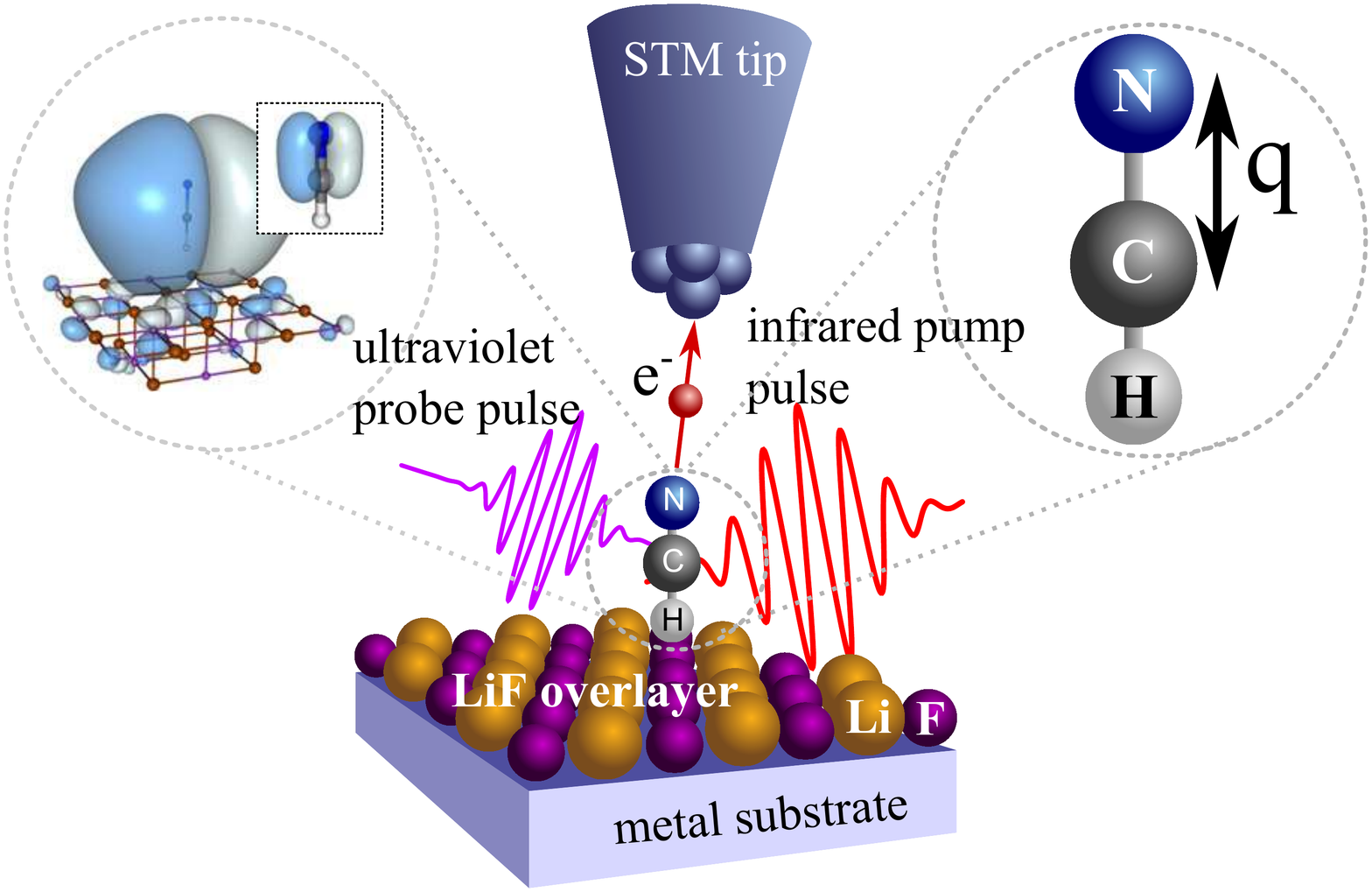}\hspace{1cm}
\caption{\small The model system proposed to investigate the local vibrational
  dynamics of a single HCN molecule adsorbed on the LiF(001) surface. After an
  infrared (IR) pump pulse has induced the vibrational dynamics, a
  second laser pulse photo-ionizes the molecule. The released electron
  collected by a STM tip can then be employed for tracing the transient
  dynamics.\label{fig1}}
\rule{\linewidth}{0.4mm}
\end{figure}

We propose in this contribution  a new model system that demonstrates the feasibility for
tracing the femtosecond dynamics of adsorbed atoms or molecules
by means of two-photon\cite{knoesel_femtosecond_1995,gruebele_femtosecond_1993,petek_surface_2002} ultrafast STM-based photoelectron detection.  As a concrete example
 (\ref{fig1}), we consider a metal substrate coated with the LiF
overlayer, where a single HCN molecule is adsorbed. The LiF layer has the
important advantage of a large band gap $E_\mathrm{g}$ (for the bulk at zero
temperature, $E_\mathrm{g}=14.2$~eV, ref.\cite{tran_accurate_2009}). Provided
the ionization energy of adsorbate is smaller than LiF work function, the
dynamics of the molecule can be accessed selectively. Furthermore, the strongly ionic character gives
rise to a particular strong bonding of the HCN molecule (which also has
a large permanent dipole moment of
1.172~a.u. \cite{thomas_balance_1993,bundgen_dipole_1995}) to the surface. Our
molecule serves as a test object representing the simplest organic molecule
and has some further convenient properties, which will be elucidated by the
analysis of the electronic properties.

Our goal is to study the transient vibrational dynamics induced by a infrared (IR)
laser pulse (the pump pulse) of moderate intensity and the vibronic coherent
motion that can
be imaged by utilizing the STM tip for recording the photocurrent (see
the (WEO1) for the video illustration). This approach is, strictly
speaking, not identical to the STM setup in the tunneling
regime. Recent experiments \cite{kirschner_photo_2013} have however
demonstrated the feasibility and the potential of the local
photoelectron detection with the tip apex, reaching a spatial
resolution in range of 5~nm.
The photoelectron is released by a second
laser pulse (the probe pulse) that ionizes the molecule and is applied
  at  a time delay $\tau$ with respect to the
pump pulse.
We remark that the laser intensities and frequencies are chosen such that
only a single electron can be released at a time.
 The delay  before  launching a second pump-probe sequence is
  large enough to allow electrons from the metal substrate to tunnel to the HCN$^+$ molecule
(characteristic time scale in the range of femtoseconds to picoseconds). On
the other hand, the molecule has to relax to the initial vibrational  state
(time scale of
picoseconds to nanosecond). Considering both of these aspects
we estimate the maximal repetition frequency of about 10~MHz for the experiment.

Based on a cluster approach for representing the LiF surface, we have found
that the
most stable equilibrium configuration of HCN is to stand upright on surface,
with the positively charged hydrogen atom directly above a F
site (further details in the Supporting Information) by bonding to the surface. The comparison the with cluster
computations reveals that the C-N and the H-C stretching
modes of the isolated HCN molecule \cite{mellau_[nu]_2010} are hardly altered
(see Supporting Information). This result is also expected from the chemical point of view due to the
strongly ionic character of both subsystems. Similarly, we  found that the highest occupied molecular orbital (HOMO) of
the HCN molecule (twice degenerated $\pi$ orbital in the isolated case) is hardly changed
by the presence of  the surface (inset in \ref{fig1}) and possesses the same symmetry.

The shape of the HOMO suggests that the excitation of the C-N
stretching mode is closely related to contracting or expanding the orbital. As
the HOMO strongly influences the photoionization properties, we expect that a
particularly clear connection between the vibrational dynamics and the
photocurrent can be established when analyzing the C-N stretching
eigenmode. We thus choose as the (one-dimensional) vibrational coordinate $q$
the distance between the carbon and the nitrogen atom (inset in \ref{fig1}). The hydrogen atom also
participates in the vibration, but with a much smaller amplitude.

\begin{figure}[ht!]
\centering
%\rule{\linewidth}{0.4mm}\vspace{0.5cm}
\includegraphics[width=8cm,angle=-90]{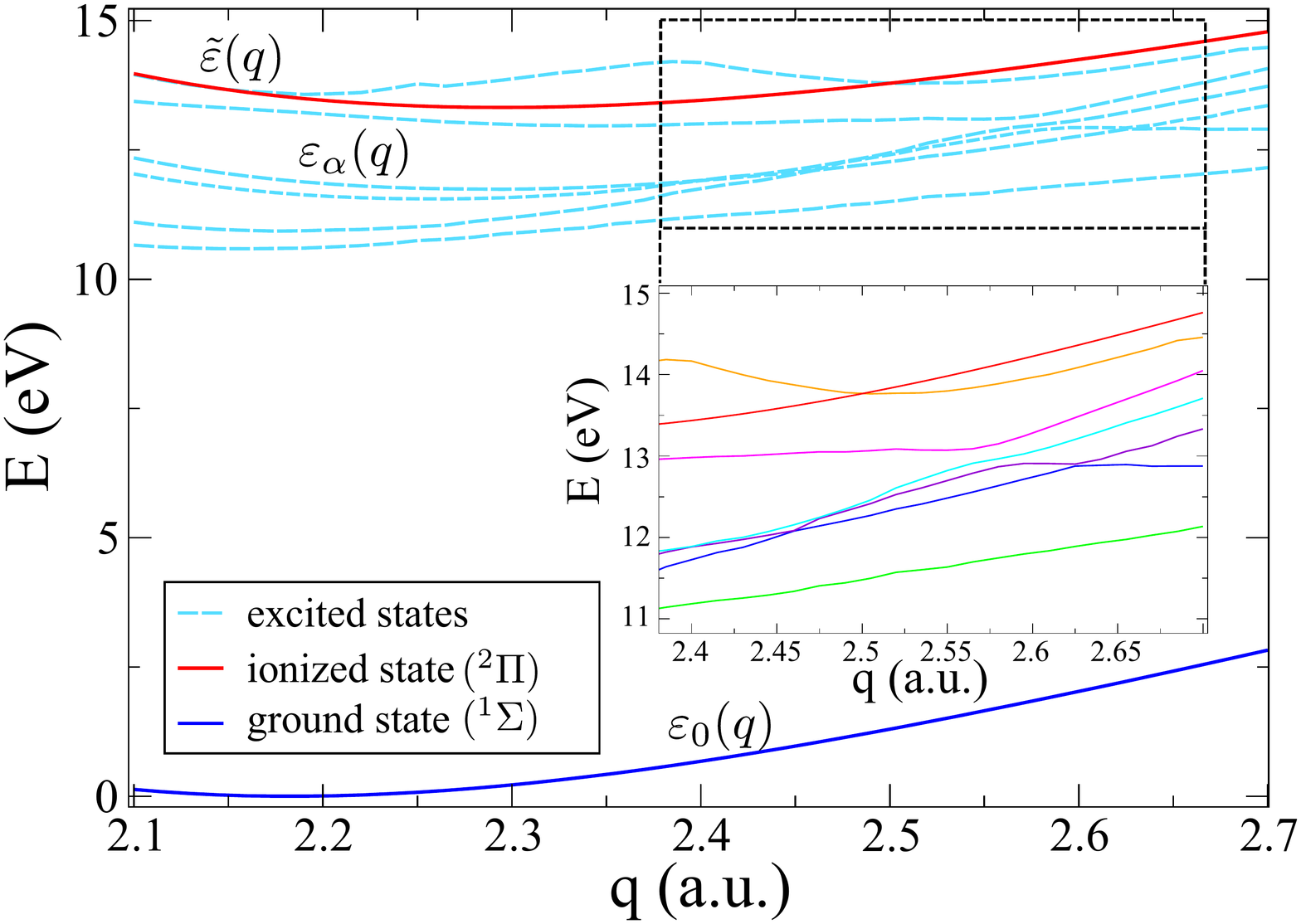}
\caption{\small The PESs of the neutral ground state (blue), the
  neutral excited states (light blue, dashed) and the ground state of HCN$^+$ (red), as a
  function of the vibrational coordinate $q$. The
  inset magnifies the region where the neutral excited states display energies
  very close to each other and shows the occurrence of avoided
  crossings. \label{fig2}}
%\rule{\linewidth}{0.4mm}
\end{figure}

For describing the laser-induced vibronic dynamics as well as the
photoionization process, we need the $q$-dependent energy
levels. We have computed all the corresponding potential energy surfaces (PESs) within
the range of 15~eV from the ground state and accounted only for those states
with a nonzero transition matrix element whilst exploiting  the dipole selection rules for
the case of a linear
polarization of the laser field set parallel to the molecule axis
(\ref{fig2}). The incident laser fields are assumed to enter under
small angle such that the components of the polarization perdendicular
to the molecule axis can be ignored. Since the molecular dynamics is slow compared to
the electronic transitions driven by the probe pulse,we can assume the bond
angle to be constant for the photoionization process and thus neglect a transition to PESs associated to a bended
conformation of the
molecule\cite{schwenzer_geometries_1974,nayak_theoretical_2005}. The computations yield the transitions pathways for the light-molecule interaction: photoionization and the electronic excitation of the neutral system (no contribution to the photocurrent). As both channels influence each other, including the neutral excited states is required for quantitative insights.

We solved the time-dependent Schr\"odinger equation (TDSE)
governing the vibrational dynamics in the presence of the infrared
laser field $E(t)$. For the latter we assume a Gaussian form $E(t)=E_0 \exp(-t^2/2
T^2_\mathrm{IR})\cos(\omega_\mathrm{IR} t)$ with $\omega_\mathrm{IR} = 329.2$~meV (this value matches
the transition energy from the ground state to the first excited state with
respect to the potential $\varepsilon_0(q)$ (\ref{fig2}). The corresponding wave length
is $\lambda_{\mathrm{IR}}=3.77$~$\mu$m) and $T_\mathrm{IR}=20$~fs. The field
amplitude derives from the intensity  \mbox{$I_{\mathrm{IR}}=1.06\times 10^{13}$
W/cm$^2$}.

\begin{figure}[ht!]
\centering
\rule{\linewidth}{0.4mm}\vspace{0.5cm}
\includegraphics[width=8cm,angle=-90]{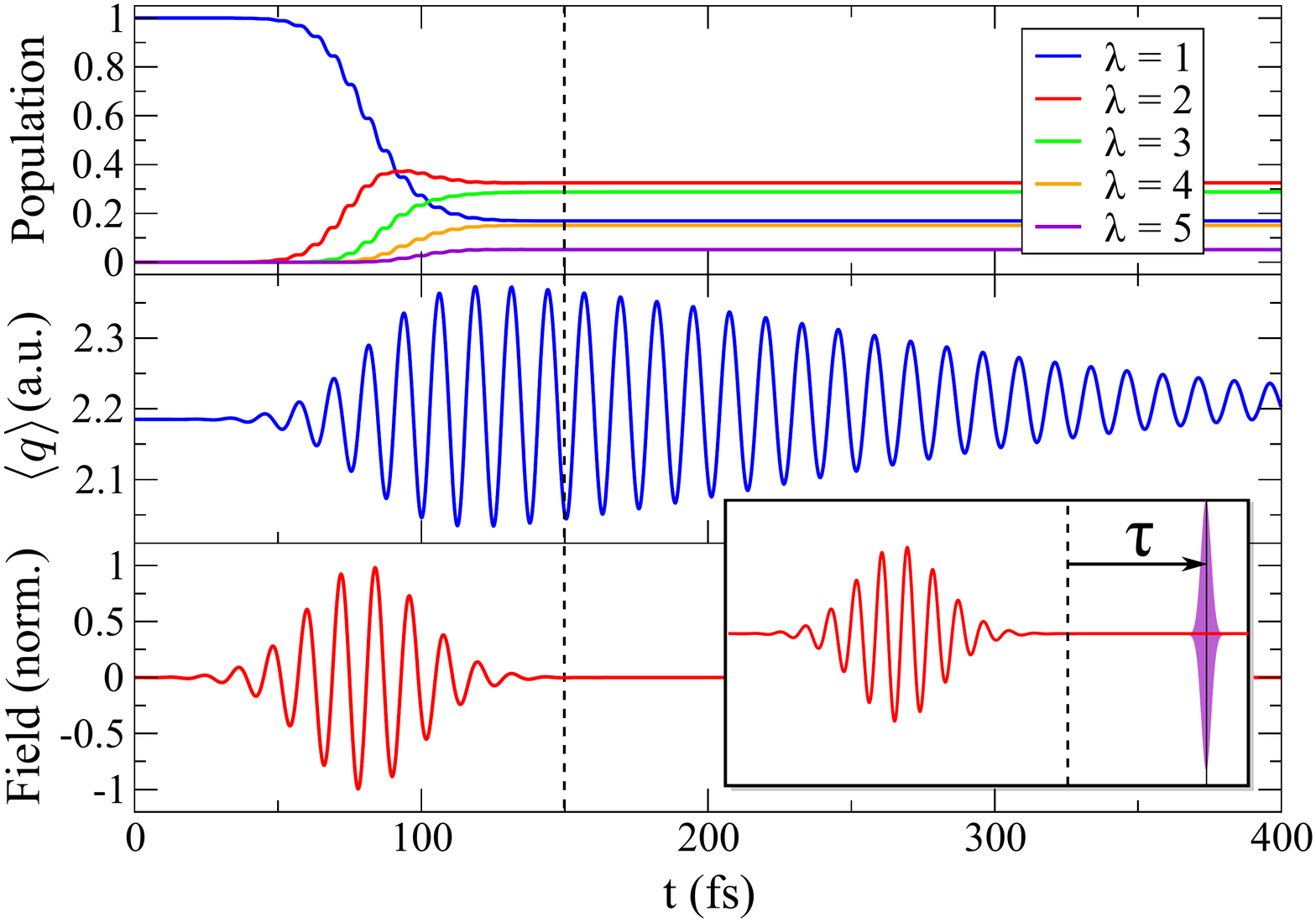}
\caption{\small
  The vibrational dynamics of the HCN molecule in the presence of the infrared
  laser field (shown in the bottom panel) in terms of the level population
  (top panel) and the time-dependent expectation value of $q$ (panel in the
  middle). The inset illustrates the definition of the pump-probe delay
  $\tau$. \label{fig3}}
\rule{\linewidth}{0.4mm}
\end{figure}

With this specific choice of the parameters the population transfer
is primarily induced from the vibronic ground state to the first excited
state, but the pulse spectral width due to the short pulse
duration      may allow
  higher
levels to participate in the dynamics (see \ref{fig3}, top panel,
where the states are labeled with the vibrational quantum number
$\lambda$). The corresponding coherent motion of the wave packet is
represented by the time-dependent expectation value of $q$ (the middle panel in
\ref{fig3}), displaying an oscillation with an amplitude of 10~\%
around the equilibrium value. The laser field $E(t)$ is shown in the bottom
panel in \ref{fig3}.

\begin{figure}[t!]
\centering
%\rule{\linewidth}{0.4mm}\vspace{0.5cm}
\includegraphics[width=4.5cm,angle=-90]{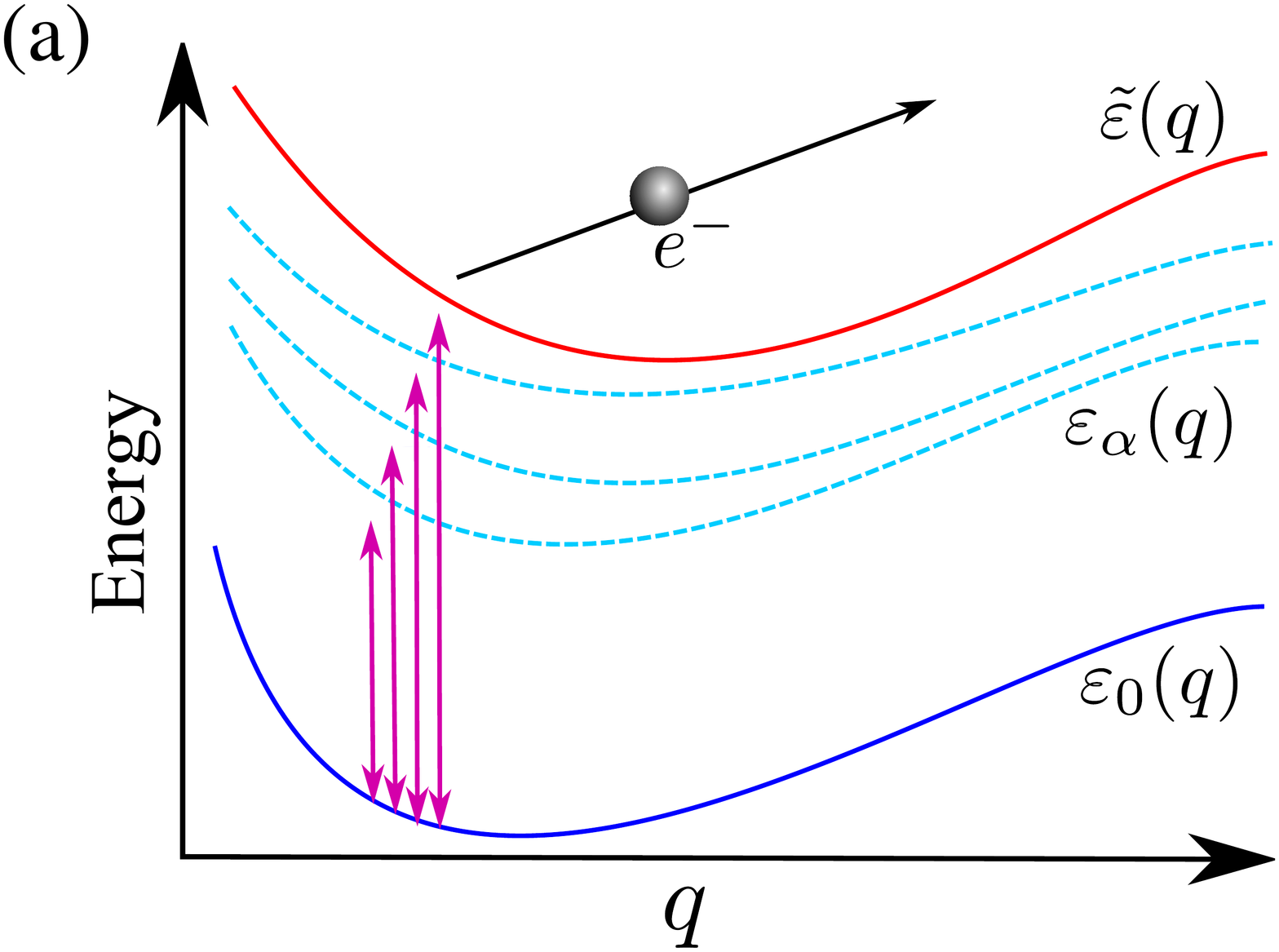}
 \includegraphics[width=4.5cm,angle=-90]{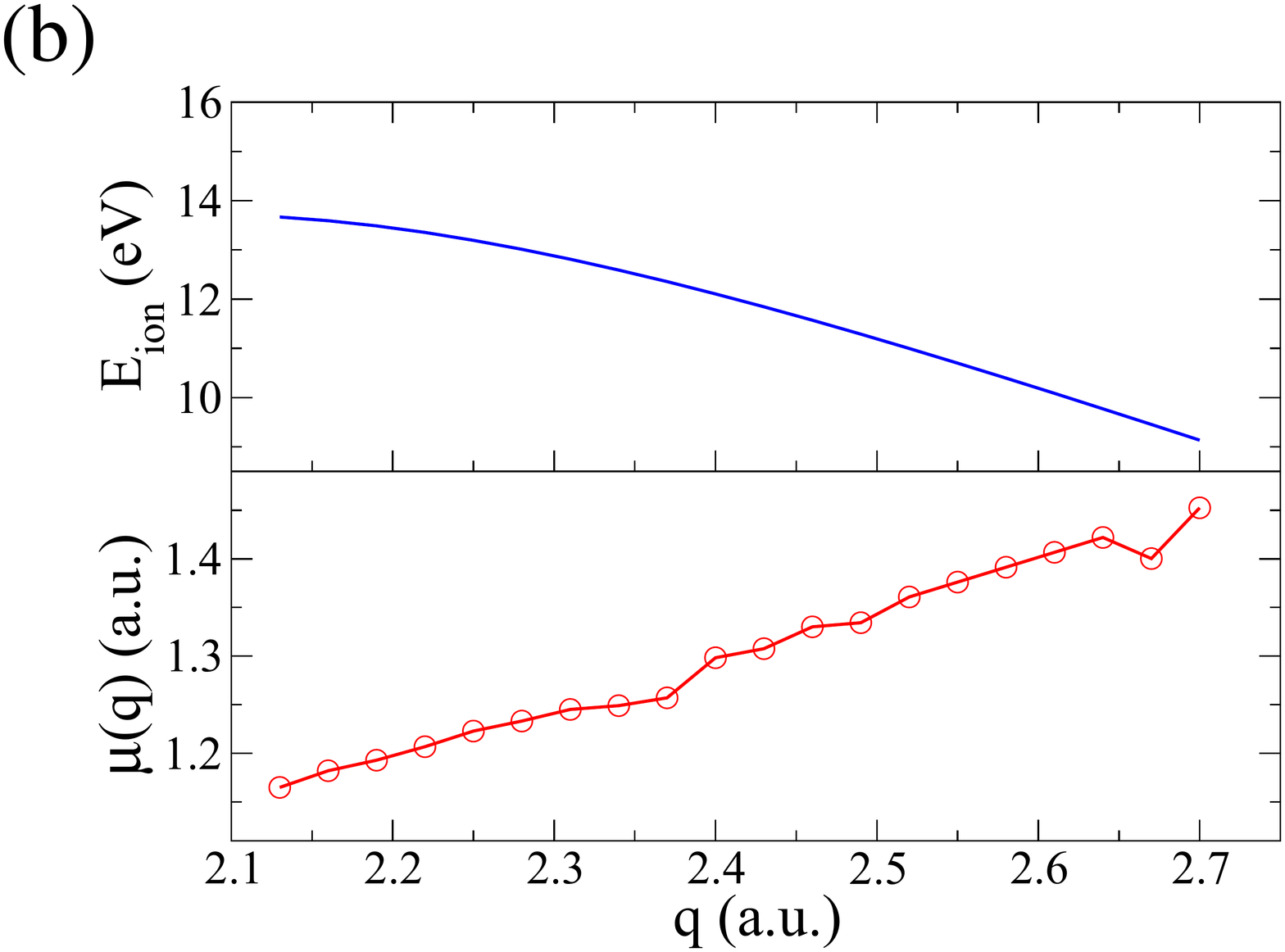}
\caption{\small (a) The simplifying model for the ionization process: the
  spectral properties of the laser field only allow for the transitions from
  the ground state to the excited states or the ionic ground state (indicated by
  the purple double arrows). (b) The $q$-dependent ionization energy and the
  momentum-integrated transition strength. \label{fig4}}
%\rule{\linewidth}{0.4mm}
\end{figure}

To unravel how the vibronic wave packet characterized in
\ref{fig3} can be traced in the quantities of interest, we analyzed the
respective many-body TDSE and derived a coupled
set of equations including the bound-state dynamics and the
release and propagation of exactly one photoelectron to the tip in the
presence of the probe pulse. The assume an ultraviolet (UV) Gaussian-shaped laser pulse (central wave length
$\lambda_{\mathrm{UV}}=100$~nm, peak intensity
\mbox{$I_{\mathrm{UV}}=3.5 \times 10^{14}$~W cm$^{-2}$} and pulse length
$T_{\mathrm{UV}}=1.97$~fs FWHH).  A detailed
derivation is provided in the Supporting Information, along with the necessary
additional approximations: (i) we ignore correlation effects of the
photoelectron with the remaining ones, (ii) ignore second-order overlap terms
of well-localized orbitals with the continuum wave function of the released electron and (iii) we employ
the Born-Oppenheimer (BO) approximation \cite{drake_springer_2005}. For the
latter, we spend a few words more on the justification for our model system. Since
the diabatic coupling matrix elements of the electronic wave functions
sandwiching the derivatives with respect to the vibrational coordinate are related to the velocity (the
kinetic energy) of the nuclei  (slow on the electronic time scale)  divided by
the energy level spacing of the states, these matrix elements play only a
minor role as long as the electronic levels are well-separated. This is,
e.~g., the case for the ground state in \ref{fig2}, but not for the
excited states. However, the physical picture of the  slow nuclei implies
that the diabatic coupling elements have only a minor influence on the
short-time dynamics, i.~e. when restricting the pulse length of the probing
electrical field to a time scale small compared to the characteristic time
scale for the vibrations, the BO approximation is still valid. This  assumption is tolerable in view of the fact that
$T_{\mathrm{UV}}/T_{\mathrm{vib}}\sim 0.1$. We furthermore specialize to  the scenario sketched in
\ref{fig4}(a), i.~e. we assume that no transitions between the neutral
excited state occur, which is ensured by the spectral properties of the probe pulse

Apart from the approximations stated above, our treatment of the coupled
population dynamics of the involved electronic states of the neutral molecule the
released electron is exact. The photoelectron wave function is represented on a
real-space grid, i.~e. it does need to be constructed invoking additional
approximations. The charge interaction of the photoelectron with the ionized
molecule and the time-dependent laser field is taken into account, as well.

The experimentally measurable quantity is  the photoionization
current. Our studies have elucidated that the key quantity connecting the vibrations in the ground state PES
$\varepsilon_0(q)$ and the photocurrent is the
transition matrix element of the Dyson orbital
with the photoelectron wave function. The Dyson orbital $\phi_0$ \cite{melania_oana_dyson_2007,oana_cross_2009}  is defined as the
overlap of the $N-$electron wave function with the $(N-1)-$electron wave
function of the ionized system. We computed $\phi_0$ by approximating the
dominant configurations of the ground states of both the neutral and the
ionized molecule by the Hartree-Fock determinants. The (expected) result
reveals that this Dyson orbital overlaps by about 90 \%
with the HOMO. Since the C-N stretching vibration directly
contracts or expands the HOMO, we expect an almost linear
dependence of the transition strength on the vibrational
coordinate. This scenario is supported by the calculation of the transition matrix
element  (integrated over the momentum of the photoelectron), depicted in
\ref{fig4}.

Solving the corresponding time-dependent Schr\"odinger equation yields the photoelectron wave
function and thus the measurable current density $\mathbf{j}$.
For the pump-probe experiment, the time-integrated current (i.~e. a
probability) rather than the actual time-dependent current $\mathbf{j}$
contains the desired information. Taking the projection in the direction of the tip yields the
spatio-resolved probability $P$ of detecting a photoelectron
in a plane parallel to the surface and entails the information
on the initial vibrational wave packet. We consider a plane
with a distance of $d=4.5$~\AA~above the nitrogen atom ($x$-$y$ plane) and compute the
detection probability as a function of the position and the time delay $\tau$  between the pump and the probe
pulse. For  $\tau=0$  the maximum of the UV pulse being
centered at $t=150$~fs (see inset in \ref{fig3}(b)).

\begin{figure}[t!]
\centering
\rule{\linewidth}{0.4mm}\vspace{0.5cm}
\includegraphics[width=5cm,angle=-90]{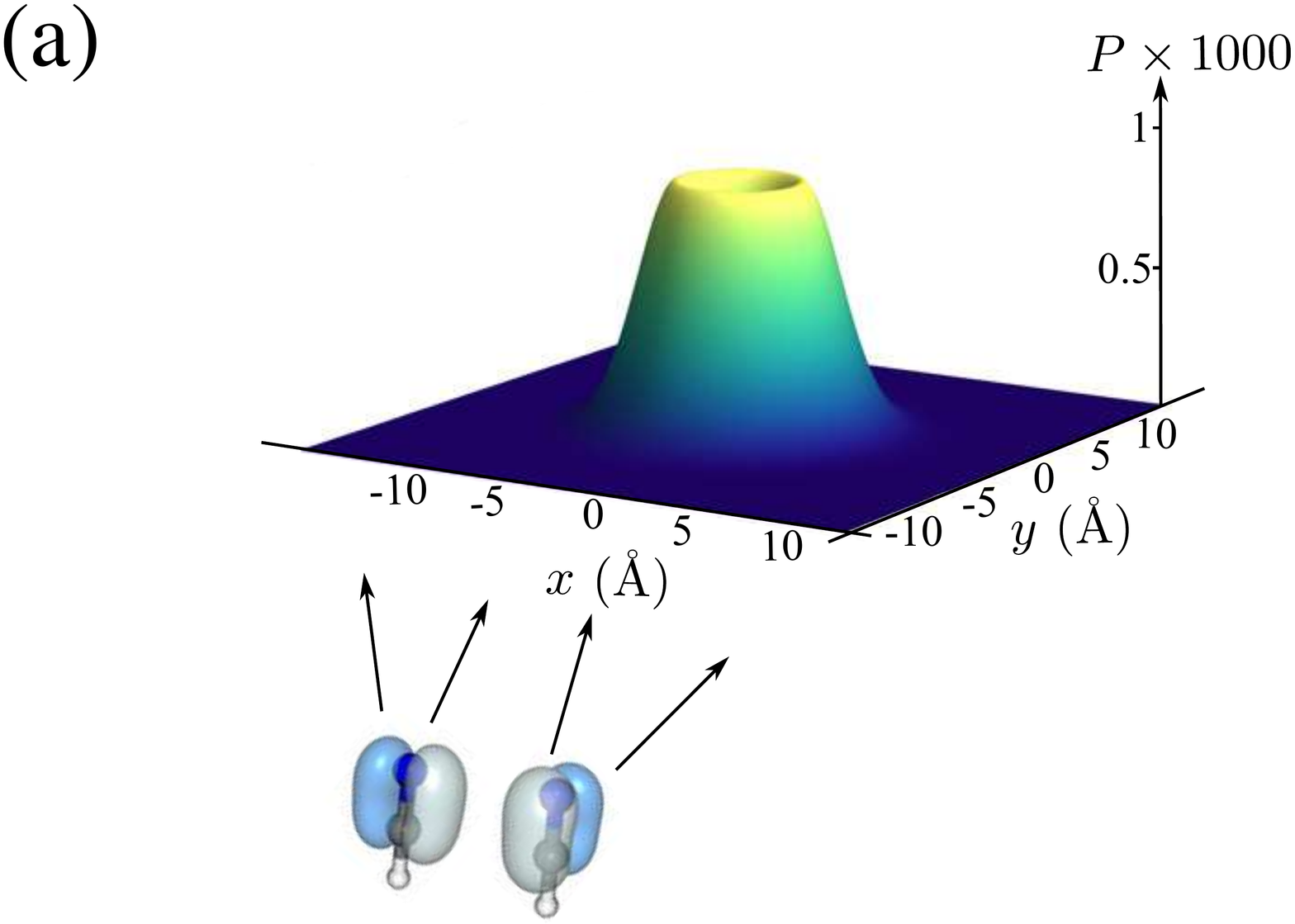} %\hspace{1cm}
\includegraphics[width=6cm,angle=-90]{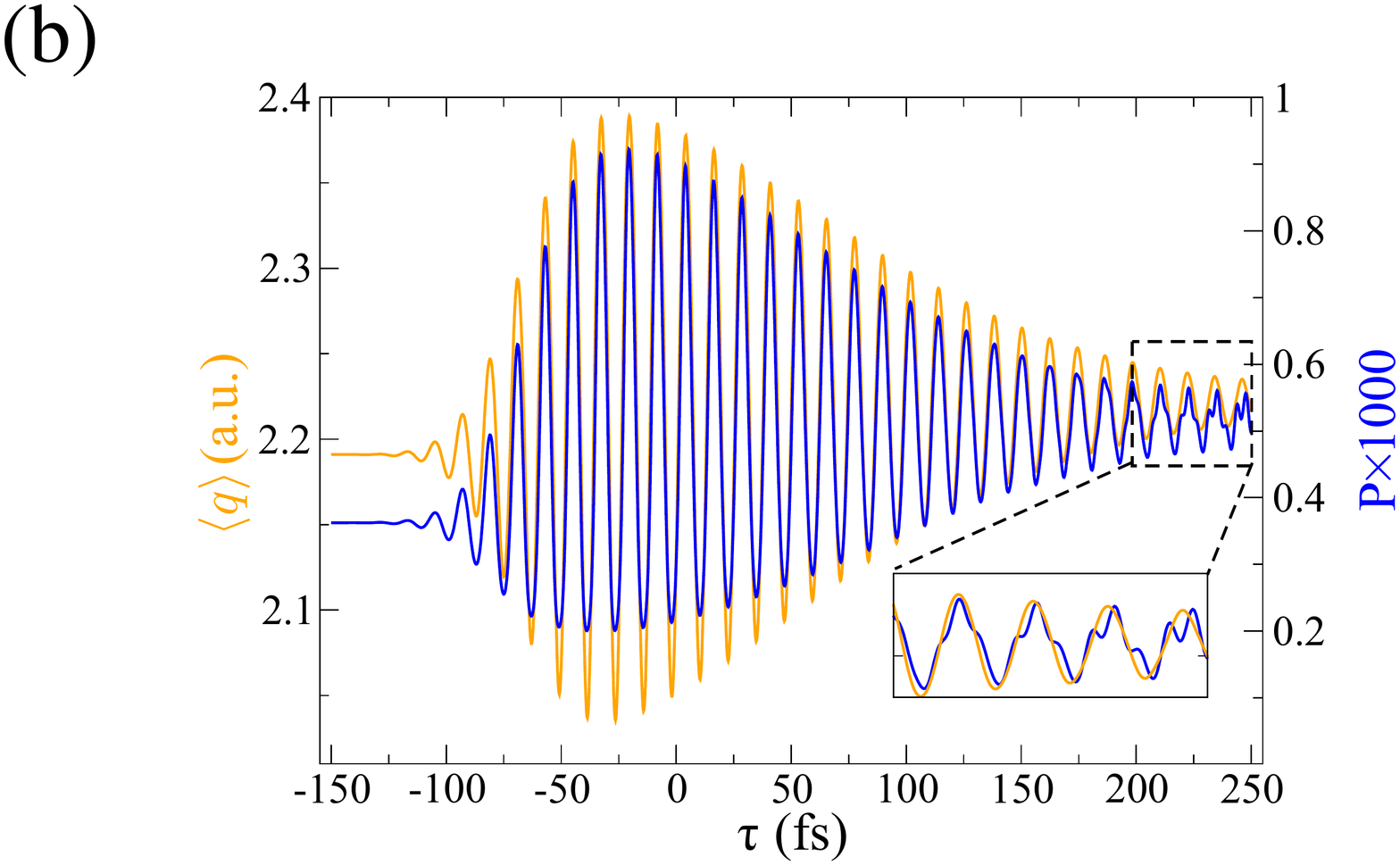}
\caption{\small (a) The spatial dependence of the detection
  probability $P$ in the shifted $x$-$y$ plane placed \mbox{$\sim 4.5$~\AA}~above the
  molecule, for fixed $\tau=-150$~fs. Both possible orientation
  directions of the degenerated Dyson orbital $\phi_0$ are taken into account.
  (b) The comparison of the expectation value $\langle q \rangle$ from
  \ref{fig3} with the detection probability $P$, as a function of the
  pump-probe delay $\tau$. The point in space for the detection is the same
  plane as in (a), such that $P$ has the maximal amplitude. \label{fig5}}
\rule{\linewidth}{0.4mm}
\end{figure}

The spatial dependence of $P$ is shown in \ref{fig5}(a). Since the
photoelectron primarily originates from the Dyson orbital $\phi_0$,
we can expect that the spatial structure of the detection probability in the
detection plane is related to a cut through the $x$-$y$ plane of $\phi_0$. At
this point, we have to take for  the degeneracy of the HOMO and add
the probability contributions according to the two orientations of the Dyson
orbital, resulting in a radially symmetric spatial dependence. Interestingly,
the probability displays a minimum directly above the molecule, as the Dyson
orbital has a nodal plane along the molecule axis. We  infer  thus that the dependence of $P$ on $x$
and $y$ yields a probability map which is closely related to viewing the Dyson
orbital ``from above''.

We proceed by fixing $x$ and $y$ such that the detection probability is
maximal and study the dependence on $\tau$. As we have already discussed, we
expect an approximately linear mapping of the coherent vibrational dynamics to
the detection probability for two reasons:  the dependence of (i) the ionization energy and
(ii) the transition strength to the continuum on $q$ is almost linear. The
result is presented in \ref{fig5}(b),
where the expectation value of the vibrational coordinate is also shown for a comparison. Both
curves are very similar though some higher-frequency components occur evidencing
 non-linear contributions.  As an outcome  of this
 study, animation (WEO2) illustrates  the
time evolution of the vibronic wave packet  simultaneously with the
$\tau$-dependent detection probability
(spatially resolved along the $x$ axis).

In conclusion, we suggested theoretically a novel experimental setup to access the spatio-temporal
dynamics of adsorbates. The proposal is based on a combination
of pump-probe techniques with a local detection scheme.
We illustrated the model by studying explicitly a sample consisting of a HCN molecule
adsorbed on  a LiF overlayer deposited on a metal substrate.
We studied how the molecule is
adsorbed on the surface and find that the vibrational as well as the electronic
properties are hardly altered. The proposed
setup involved (i) the excitation of a coherent vibronic wave
packet due to an infrared pump pulse, and (ii) the photoionization of the adsorbed molecule
by an ultraviolet probe pulse.  (iii) The photoelectrons are detected by the STM tip.
We  demonstrated by a proof-of-principle calculation how the vibronic wave
packet  can be mapped onto a  probability $P$ detected by the STM tip.
The tip position yields
the spatial dependence and can be exploited to image  the involved
orbitals. The temporal dependence with respect to the pump-probe delay
$\tau$ yields a measurable signal which closely resembles the time-dependent
expectation value of the vibrational coordinate $q$, characterizing the
coherent dynamics.
While our proof-of-principle study concentrates on a rather simple
molecule, a generalization to more complex systems is conceptionally
straight forward. The combination of spatial and temporal resolution
potentially coalesces in, e.~g., probing relaxation and decoherence processes or
photo-induced conformation
switching on single molecules in the time domain.

\section{Theoretical methods}
All structure computations have been carried out using the \textsc{Gaussian}
03 code. Details (method, basis set) for each of the steps are provided in the
Supporting Information. For the time-propagation of the photoelectron wave
function we employed a fourth-order Runge-Kutta scheme and discretized the
spatial derivatives to the fourth order.

\begin{acknowledgement}
This research is supported financially by the DFG through  SFB 762.
\end{acknowledgement}

\section{Associated content}

\textbf{Web Enhanced Object 1} (WEO1). Schematic video of the model system and the
setup. Available in the HTML version of the paper.\\
\textbf{Web Enhanced Object 2} (WEO2). Animation of the time-delay dependence of the
vibrational wave-packet dynamics synchronized with the detection
probability. Available in the HTML version of the paper.\\
\textbf{Supporting Information Availabe:}. Details for the  computations of the
vibrational modes and the orbitals of the surface-molecule system,
for calculating the PESs and the derivation of Schr\"odinger equations for the
photoemission. Available free of charge via the Internet at \mbox{\url{http://pubs.acs.org}}.

%%%%%%%%%%%%%%%%%%%%%%% References %%%%%%%%%%%%%%%%%%%%%%%%%

%%% decreasing global font size %%%
\small

\providecommand*\mcitethebibliography{\thebibliography}
\csname @ifundefined\endcsname{endmcitethebibliography}
  {\let\endmcitethebibliography\endthebibliography}{}

\end{document}